\documentclass[%
 amsmath,amssymb,
 aps,
]{revtex4-2}

\usepackage[english]{babel}

\usepackage{tikz}
\usetikzlibrary{angles,quotes}
\usetikzlibrary{calc}
\usetikzlibrary{patterns}

\usepackage{xcolor}
\usepackage{verbatim}

\usepackage{graphicx}% Include figure files
\usepackage{dcolumn}% Align table columns on decimal point
\usepackage{bm}% bold math
\pdfoutput=1
\usepackage{url}
\usepackage{hyperref}
\usepackage{amsfonts}% Include figure files
\usepackage{epsfig}
\usepackage{dcolumn}% Align table columns on decimal point
\usepackage{xcolor}
\usepackage{amsmath}
\usepackage{amssymb}
\usepackage{amsthm}
\usepackage{amsfonts}
\usepackage{verbatim}
\usepackage{enumerate}
\usepackage{graphicx}
\usepackage{accents}
\usepackage{bm}% bold math

\usepackage{physics}
\usepackage{tensor}
\usepackage{siunitx}
\usepackage{circuitikz}
\usepackage{wasysym}

\usepackage{graphics}
\usepackage{caption}
\usepackage{subcaption}
\usepackage{epstopdf}
\usepackage[utf8]{inputenc} % set input encoding (not needed with XeLaTeX)
\usepackage{pgfplots}
\usepackage{slashed}

\usepackage{braket}
\usepackage{dsfont}
\usepackage{float}

\newcommand{\vnur}{\bar{\nu}}
\newcommand{\cL}{\mathcal{L}}

\newcommand{\ed}[1]{}

\begin{document}

\title{Sterile Neutrino Dark Matter from\\ Generalized $CPT$-Symmetric Early-Universe Cosmologies}

\author{Adam Duran}
\email{adsduran@ucsc.edu}
\author{Logan Morrison}
\email{loanmorr@ucsc.edu}
\author{Stefano Profumo}
\email{profumo@ucsc.edu}

\affiliation{Department of Physics and Santa Cruz Institute for Particle Physics\\University of California, Santa Cruz, CA 95064, USA}

\begin{abstract} We generalize gravitational particle production in a radiation-dominated $CPT$-symmetric universe to non-standard, but also $CPT$-symmetric early universe cosmologies. We calculate the mass of a  right-handed ``sterile'' neutrino needed for it to be the cosmological dark matter. Since generically sterile neutrinos mix with the Standard Model active neutrinos, we use state-of-the-art tools to compute the expected spectrum of gamma rays and high-energy active neutrinos from ultra-heavy sterile neutrino dark matter decay. We demonstrate that the sterile neutrinos are never in thermal equilibrium in the early universe. We show that very high-energy Cherenkov telescopes might detect a signal for sterile neutrino lifetimes up to around 10$^{27}$ s, while a signal in high-energy neutrino telescopes such as IceCube could be detectable for lifetimes up to 10$^{30}$ s, offering a better chance of detection across a vast landscape of possible masses.
\end{abstract}

\maketitle

\section{Introduction}
Dark matter (DM) continues to stand as one of the greatest mysteries at the interface of particle physics and cosmology \cite{Zyla:2020zbs}. Since no particle in the standard model (SM) can account for the bulk of the cosmological dark matter, one must extend the particle content of the theory to include at least an additional particle species, or a bound state thereof, that can account for a matter species making up around five times the baryonic matter. Adding right-handed neutrino (RHN) states is a well-motivated and oft-followed route that can at once address the origin of active neutrino masses, explain the matter-antimatter asymmetry, and provide a viable, arguably minimal, DM candidate \cite{Boyarsky:2018tvu}. 

Recently, Ref.~\cite{Boyle:2018rgh, Boyle:2018tzc} considered a peculiar and interesting possibility for the generation of the cosmological dark matter: while in flat space-time the ``natural'' vacuum is the one respecting the isometries of Minkowski space, in a general, curved space-time this is no longer true. The choice of vacuum is generally observer-dependent (for instance to inertial observers at different points in the space-time), and different observers will define different, inequivalent vacua. As a result, the zero-particle state according to one observer can actually be populated by particles according to a different observer (this is well known in the context of Hawking radiation from black hole evaporation, or the Unruh radiation experienced by an accelerated observer) \cite{Birrell:1982ix, Mukhanov:2007zz}. In particular, this holds generally for Friedmann-Robertson-Walker (FRW) cosmologies. Ref.~\cite{Boyle:2018rgh, Boyle:2018tzc} realized that assuming time-reversal invariance allows to identify uniquely a $CPT$ invariant vacuum in a FRW background, enabling a well-defined calculation of the number density of gravitationally produced particles as seen by a late-time observer.

Gravitational particle production in an FRW background without inflation, similar to the scenario under consideration here (albeit, here without the additional requirement of $CPT$ invariance), has been studied extensively in the past (for reviews, we refer the Reader to Ref.~\cite{DeWitt:1975ys, Lawrence:1995ct}).  The key theoretical results are that particle production (specifically fermions, or conformally coupled scalars) occurs at the cosmological epoch when
particle masses are comparable to the Hubble expansion rate; the particle number density $n\sim m^3$, and particles are created with the equation of state of dust; finally, the fractional relic density of these particles $X$ at the time of radiation-matter equality was found to be $\Omega_X\sim(m_X/10^9\ {\rm GeV})^{5/2}$ \cite{Kuzmin:1999zk}.  Further studies considered similar scenarios, but in the presence of inflation (see e.g. \cite{Chung:2011ck} and references therein).

If the early universe can be described by a conformally flat FRW metric, the early-time and the late-time (or the 'in' and 'out') vacuum states are not equivalent due to the lack of time-translation symmetry in the metric. Calculating the number density of particles then involves performing a Bogoliubov transformation on the quantum field modes and finding the appropriate Bogoliubov coefficients. Remarkably, Ref.~\cite{Boyle:2018rgh} found that the standard in/out vacuum states do not respect $CPT$, and constructing a $CPT$ invariant vacuum state leads to a certain unique form of the relevant Bogoliubov coefficient \cite{Boyle:2018rgh,Boyle:2018tzc}. Given the number density produced by this expansion mechanism, if this species were to account for the total dark matter (DM) density $\rho_{DM}$, then its mass would have to be around 4.8$\times$10$^8$ GeV. This calculation was performed in the context of an early universe cosmology described by a radiation-dominated cosmology, with a scale factor $a(\tau)\propto \tau$, where $\tau$ indicates the conformal time, and for the most natural and minimal DM candidate, the right-handed neutrino, assumed to be absolutely stable.

Here, we note that the requirement of $CPT$ symmetry is not unique to radiation domination: as we explain in this work, the universe could have been dominated, at early times, by some other species with a different energy density, and yet satisfy $CPT$ symmetry. Note that the gravitational production of particles discussed here does not touch upon the question of whether the issues of homogeneity and isotropy are addressed or not, and only relies on $CPT$ symmetry at positive conformal times \cite{Boyle:2018rgh, Boyle:2018tzc}.

The possibility of an alternate cosmological history prior to Bug Bang Nucleosynthesis has received considerable attention (for a recent review see Ref.~\cite{Allahverdi:2020bys}), and here, as far as our motivation, we build largely on the literature on the topic. The key observational fact for radiation domination is the synthesis of light elements; prior to that, there is no observational evidence that underpins radiation domination. In fact, the possibility of an early matter-dominated phase is very well justified and has been widely considered \cite{Barenboim:2013gya}; other causes of nonstandard early-universe expansion history include single or multi-field post-inflation reheating , multi-step or thermal inflation (in fact the presence of any extra stages of inflation, see e.g. Ref.~\cite{Bassett:2005xm} for a review), heavy particles and dark sectors \cite{Chung:1998rq}, moduli fields \cite{Acharya:2009zt}; the resulting structure of the cosmological expansion history impacts, among other things, gravitational particle production, as was noted in a recent study \cite{Hashiba:2019mzm} that considered the case of quintessential cosmology in the early universe, with a formalism not dissimilar to the one employed here. Note, however, that the present setup is uniquely different from previous work where gravitational production is connected to inflation.

A non-standard FRW background leads to different equations of motion for the right-handed neutrino field, which, we note here, can be approximated using the WKB method. Once the appropriate Bogoliubov coefficient is computed from the WKB solutions in the asymptotic future, one can follow a similar analysis to Ref.~\cite{Boyle:2018rgh} and calculate the resulting particle number density at late times. The final result of this exercise is a DM mass as a function of a parameter which indicates the scaling of the energy density with the cosmological scale factor and, as we discuss below, a second parameter that depends on when the universe transitions from the non-standard early-times cosmology to radiation domination, needed from cosmological observations to occur prior to the synthesis of light elements \cite{Zyla:2020zbs}. 

%We note that gravitational particle production in the early universe has been extensively discussed in the different setting of inflationary cosmology (see e.g. Ref.~\cite{Chung:2011ck}), including in the case of quintessential cosmology \cite{Hashiba:2019mzm}, with a formalism not dissimilar to the one employed here. Note, however, that the present setup is uniquely different from previous work where gravitational production is connected to inflation.

In this work, besides generalizing the results of \cite{Boyle:2018rgh,Boyle:2018tzc} to any $CPT$-symmetric early-universe cosmology, we relax a key assumption in the dark matter sector: that the right-handed neutrino dark matter be absolutely stable because of a discrete symmetry. Relaxing such assumption produces the interesting prospect of testing experimentally this production mechanism by searching for the decay products of massive right handed neutrinos, with a lifetime related to the mixing angle between sterile and active neutrinos. In turn, this calls for addressing the question of whether the gravitationally-produced neutrinos can ever thermalize, which we address in detail in sec.~\ref{sec:thermalization}.

The remainder of this work is structured as follows: in the following section \ref{sec:dmnumberdensity} we describe in detail the calculation of the dark matter number density generated in arbitrary $CPT$ symmetric cosmologies in the early universe; the following sec.~\ref{sec:decay} discusses the right-handed neutrino sector and the decay modes thereof; sec.~\ref{sec:thermalization} explores the issue of thermalization of right-handed neutrinos; sec.~\ref{sec:lifetime} addresses constraints and prospects for the detection of decaying massive right-handed neutrinos as predicted in the present scenario; finally, sec.~\ref{sec:conclusions} concludes.

\section{Sterile Neutrino as Dark Matter in Generalized $CPT$ Symmetric Cosmologies}\label{sec:dmnumberdensity}

Consider a cosmology dominated by a species with equation of state $P=w\rho$, and thus with an energy density-scale factor relation
\begin{equation}
    \rho\sim a^{-3(1+w)}.
\end{equation}
Recalling that, for $w\neq1$, time $t\sim a^{3(1+w)/2}$, and that conformal time is defined as
\begin{equation}
    \tau\equiv\int\frac{dt}{a},
\end{equation}
the relation between conformal time and scale factor for an equation of state parameter $w$ is
\begin{equation}
    \tau\sim a^{\frac{1+3w}{2}},\quad {\rm or}\quad a(\tau)\sim \tau^{\frac{2}{1+3w}}.
\end{equation}
The requirement for $CPT$ symmetry is that $a(\tau)=-a(-\tau)$; this, in turn, requires
\begin{equation}
    \frac{2}{1+3w}={\rm odd}=N=2k+1,\ k=0,\pm1,\pm2,\dots
\end{equation}
As a result, we obtain the equation of state of radiation ($w=1/3$) for $N=1$, $-1/3<w<-1/9$ equations of state for positive $N\neq2$ (we disregard the $w=0$ case), and $-1<w<-1/3$ for negative $N$. We do not concern ourselves with the issue of, and the specific model for, how the transition between the era when this species dominates in the early universe and radiation domination occurs, since this does not yield any effect on what we concern ourselves with here, i.e. gravitational particle production. 

    We hereafter generalize the particle production in a radiation-dominated $CPT$-symmetric universe discussed in  Ref. \cite{Boyle:2018tzc,Boyle:2018rgh} to the non-standard cosmologies just described above. 
    Ref.~\cite{Boyle:2018tzc,Boyle:2018rgh} considered a heavy sterile neutrino essentially decoupled from the thermal bath of the early universe and constructed the corresponding quantum field coupled to the expanding Friedmann-Robertson-Walker (FRW) metric 
    \begin{equation}
        ds^2=a^2(\tau)[-d\tau^2+dx^2]
    \end{equation}
    that is asymptotically flat at early ($\tau\rightarrow-\infty$) and late ($\tau\rightarrow+\infty$) times. The Lagrangian for the spinor field in the expanding background is given by the usual
\begin{equation}
          {\cal L}\supset i\bar\psi\slashed{\partial}\psi-\mu\bar\psi\psi,
\end{equation}
    with equation of motion
    \begin{equation}\label{eq:motion}
        (i\slashed{\partial}-\mu)\psi=0.
    \end{equation}
    Following \cite{Boyle:2018rgh}, we posit that the right handed neutrino  mass term originates from the vacuum expectation value of a field $\Phi$ whose equation of motion at early times, prior to the electroweak phase transition, is simply
    \begin{equation}
        \frac{1}{2}\left(\frac{d\Phi}{d\tau}\right)^2=a^4\rho=\sim a^4a^{-3(1+w)}\sim \tau^{N(1-3w)}=\tau^{2N-2}.
    \end{equation}
   We call $\tau_1$ the conformal time at which the universe switches from being dominated by a non-standard energy density $\rho\sim a^{-3(1+w)}\sim a^{-2\frac{N+1}{N}}\sim \tau^{-2(N+1)}$ to radiation domination. Solving the equation above gives
    \begin{align}
        \dv{\Phi}{\tau} = \sqrt{2\rho}a^2 \sim \tau^{N-1} = \begin{cases}
            C_{1}\tau^{N-1} & 0 < \tau < \tau_1\\
            C_{2} & \tau > \tau_1\\
        \end{cases}
    \end{align}
    and thus:
    \begin{align}
        \Phi(\tau) = \begin{cases}
            C_{1}\frac{\tau^N}{N}, & 0 < \tau < \tau_1\\
            C_{1}\frac{\tau_1^N}{N} + C_{2}\qty(\tau-\tau_1), & \tau > \tau_1\\
        \end{cases}
    \end{align}
    and hence, the effective mass is:
    \begin{align}
        \mu(\tau) = y\Phi(\tau) = y\times\begin{cases}
            C_{1}\frac{\tau^N}{N}, & 0 < \tau < \tau_1\\
            C_{1}\frac{\tau_1^N}{N} + C_{2}\qty(\tau-\tau_1), & \tau > \tau_1\\
        \end{cases}
    \end{align}
   
   To find the constants in the equation above, first ask that the asymptotic value in the large $\tau$ limit give the ``right'' radiation domination answer, i.e. $\mu(\tau)=y(2\rho_1)^{1/2}\tau=\gamma\tau$, with $\rho_1$ the value at the ``beginning'' of radiation domination, i.e. for us at conformal time $\tau_1$, and $y=M/\hat\mu$: this sets $C_2=1$; second, demand continuity of both $\Phi(\tau)$ (no constraints) and of its derivative, which imposes $C_1\tau_1^{N-1}=C_2=1$ thus $C_1=1/\tau_1^{N-1}$. So indeed for large $\tau<\tau_1$, where we hypothesize production occurs, $\mu(\tau)=\gamma\tau^N/\tau_1^{N-1}$. 
    
    Given the symmetries associated with the FRW background, different observers disagree on what is a positive- or a negative-frequency mode: each observer will have their own corresponding mode functions, operators, and vacuum states. Ref.~\cite{Boyle:2018tzc,Boyle:2018rgh} label these the ingoing `$-$' and outgoing `+' states, for observers in the asymptotic, conformally invariant past, and in the non-conformally invariant late-time universe. The fields expansion for these modes are given by 
    \begin{equation}
        \psi(x)=\int\frac{d^3p}{(2\pi)^3}\bigg[a_{\pm}(p,h)u_{\pm}(p,h,\tau)e^{ipx}+b_{\pm}^\dagger(p,h)\upsilon_{\pm}(p,h,\tau)e^{-ipx}\bigg].
    \end{equation}
    In the equation above, the operators $a_{\pm}(p,h)$ and $b_{\pm}(p,h)$ act on their respective vacuum states of the asymptotic observers, which are defined by
\begin{gather}
a_-(p,h)\ket{0_-}=b_-(p,h)\ket{0_-}=0,
\\
a_+(p,h)\ket{0_+}=b_+(p,h)\ket{0_+}=0.
\end{gather}
    In order to relate the field expansions and operators of the two asymptotic observers, one employs the Bogoliubov transformation. The key result of this transformation is that mode functions and operators for  early-time observers can be expressed as a linear combination of the late-time observer's mode functions and operators, which form a complete set of states, as
    \begin{gather}
        u_-(p,h,\tau)=\alpha(p,h)u_+(p,h,\tau)+\beta(p,h)\upsilon_+(p,h,\tau),
        \\
         a_-(p,h)=\alpha(p,h)a_+(p,h)+\beta(p,h)a_+^\dagger(p,h),
    \end{gather}
    where $\alpha(p,h)$ and $\beta(p,h)$ are the so-called Bogoliubov coefficients. If we apply the number operator of the ingoing observer $N_-=a^\dagger_-a_-$ to the outgoing observers vacuum state, the result is 
    \begin{equation}
        \braket{0_+|N_-|0_+}=|\beta(p,h)|^2.
    \end{equation}
    Integrating over the phase space gives the mean particle number density  at late times, one gets the desired result
    \begin{equation}
      n=2\sum_{h}\int\frac{d^3p}{(2\pi)^3}\mid\beta(p,h)\mid^2.  
    \end{equation}
    The equation above implies that particle number is observer-dependent, which is a well-known fact in the context of the theory of quantum fields in expanding backgrounds. However, Ref.~\cite{Boyle:2018tzc,Boyle:2018rgh} found that since the ingoing and outgoing vacuum states are not equivalent, they are not $CPT$ invariant (See section 3.3 in \cite{Boyle:2018rgh}). To correct this, Ref.~\cite{Boyle:2018tzc,Boyle:2018rgh} constructed a $CPT$ invariant vacuum state that respects the transformations of the field. A Bogoliubov transformation matrix considering $CPT$ transformations was calculated along with the corresponding coefficients (See section 3.4 in \cite{Boyle:2018rgh}). The coefficient of interest $\beta_+(p,h)$ was calculated to be
    \begin{equation}\label{eq:beta}
         \beta_+(p,h)=\sin{\bigg[\frac{1}{2}\arcsin{\big(\beta(p,h)\big)}\bigg]},
    \end{equation}
    where this quantity gives the mean particle density given the assumed $CPT$ invariant vacuum state of the early universe. The Bogoliubov coefficient $\beta(p)$ in the inverse sine is calculated considering the ingoing and outgoing vacuum states. 
    
    In the context of the generalized conformally invariant cosmologies we consider here, the Bogoliubov coefficient for the ingoing and outgoing states is corrected by an effective mass that now reads $\mu(\tau)=\gamma\tau^N/\tau_1^{N-1}$.
Since the fields mode functions form an orthonormal basis, one can follow the normalization condition
    \begin{equation}\label{eq:id}
        \sum_{h}\big[u_{\pm}(p,h,\tau)u_{\pm}^*(p,h,\tau)+\upsilon_{\pm}(p,h,\tau)\upsilon_{\pm}^*(p,h,\tau)\big]=\mathds{I}_{\textrm{4x4}};
    \end{equation}
     the Bogoliubov coefficient $\beta(p)$ in (\ref{eq:beta}) can be then calculated using the inner product of ingoing and outgoing mode functions,
    \begin{equation}\label{eq:inout}
        \braket{\upsilon_{-}|u_{+}}=\sum_{h}\big[\upsilon_-^*(p,h,\tau)u_{+}(p,h,\tau)\big]=\beta(p).
    \end{equation}
    The next task is to solve the equation of motion (\ref{eq:motion}) and find the asymptotic form. 
    For a free Weyl invariant scalar field defined on an expanding FRW metric with an effective mass $\mu$ the Lagrangian is given by
\begin{align}
    L=(\varphi')^\dagger(\varphi')-(\nabla\varphi)^\dagger(\nabla\varphi)-\mu^2\varphi^\dagger\varphi 
\end{align}
with the equation of motion
\begin{equation}
        (\partial_\tau^2-\nabla^2+\mu^2)\varphi=0.   
\end{equation}
Taking the solutions to momentum space: $\varphi(x)=u(p,\tau)e^{ipx}$ turns the equation of motion into that of a harmonic oscillator equation with frequency $\omega(\tau)$,
\begin{align}
     u''(p,\tau)+\omega^2(\tau)u(p,\tau)=0\quad;\quad\omega^2(\tau)=p^2+\gamma^2\tau^{2N}/\tau_1^{2N-2}.
\end{align}
To see how the mode solutions vary with conformal time $\tau$, we numerically solved the equation for various $N$ values, and compared with an approximate solution from the WKB formalism \cite{Feynman:1948ur}. For the ingoing and outgoing observers the WKB solutions are given by
\begin{equation}
            u_\pm(p,\tau)=\frac{1}{\sqrt{\pm 2\omega(p,\tau)}}\exp\bigg[\mp i \int_{\pm \tau_0}^{\tau}\omega(p,\tau^{'}) d\tau^{'}  \bigg]  
\end{equation}
where $\tau_0$ is arbitrary and can be fixed for convenience. The integral can be carried out analytically, and the result is given by
\begin{equation}\label{eq:uglybeast}
   \int ... =\frac{\tau}{N+1}\bigg(\sqrt{p^2+\gamma^2\tau^{2N}/\tau_1^{2N-2}}+\\
      pN_2F_1\bigg(\frac{1}{2},\frac{1}{2N},1+\frac{1}{2N};-\frac{\gamma^2\tau^{2N}/\tau_1^{2N-2}}{p^2}\bigg)\bigg)\bigg\vert_{\pm\tau_0}^\tau .
\end{equation}
This expression simplifies considerably in the large $\tau$ limit. Here, we assume that particle production occurs at conformal times smaller than the time $\tau_1$ at which the universe becomes radiation dominated. Thus, the large $\tau$ limit here means that $\tau_1 y T_1\gg 1$, i.e. $\tau_1 T_1\gg\frac{1}{y}=\frac{M_P}{M}$, with $M$ the right-handed neutrino mass. 

Using
$$
\frac{T}{T_{\rm BBN}}=\frac{a_{\rm BBN}}{a}\longrightarrow T=\frac{T_{\rm BBN}a_{\rm BBN}}{a},
$$
we have
$$
\tau_1 T_1=\frac{T_{\rm BBN}a_{\rm BBN}}{a_1}\int_0^{a_1}\frac{da}{a^2 H(a)},
$$
where $H(a)\simeq H_0\sqrt{\Omega_w}a^{-(N+1)/N}$ at $a<a_1$ by assumption, where $\Omega_w$ is the relative energy density of the exotic component,
$$
\Omega_r a_1^{-4}=\Omega_w a^{-(2N+2)/N}\longrightarrow \Omega_w=\Omega_r a_1^{(2-2N)/N}.
$$
We thus find
$$
\tau_1 T_1=\frac{T_{\rm BBN}a_{\rm BBN}}{a_1H_0\sqrt{\Omega_r}a_1^{-(N+1)/N}}\int_0^{a_1}\frac{da}{a^2 a^{-(N+1)/N}}=\frac{T_{\rm BBN}a_{\rm BBN}}{H_0\sqrt{\Omega_r}a_1^{1/N}}\left(N a_1^{1/N}\right).
$$
Finally, we find that the condition {\em does not depend on $a_1$}, and is always satisfied since, putting in the numbers (with $H_0\sim 10^{-42}$ GeV and $\Omega_r\sim 5\times 10^{-5}$), we have
$$
\frac{M}{M_P}\gg\frac{H_0\sqrt{\Omega_r}}{N T_{\rm BBN}a_{\rm BBN}N}\simeq\frac{4\times 10^{-32}}{N},
$$
implying $M\gg 10^{-5}$ eV. In order to be the dark matter, the RHN must be heavier than at least a few keV, so the large-$\tau$ condition is always satisfied.

In the large $\tau$ limit, the first term in Eq.~(\ref{eq:uglybeast}) reads
\begin{equation}
\sqrt{p^2+\gamma^2\tau^{2N}/\tau_1^{2N-2}}\approx \gamma\tau^N/\tau_1^{N-1}.
\end{equation}
Upon integration, one gets for that first term:
\begin{equation}
\frac{\gamma}{N+1}(\tau^{N+1}/\tau_1^{N-1}-(\pm\tau_0)^{N+1}/\tau_1^{N-1}).
\end{equation}
For the hypergeometric function in Eq.~(\ref{eq:uglybeast}), in the same limit, one can make the replacement 
\[
_2F_1\bigg(\frac{1}{2},\frac{1}{2N},1+\frac{1}{2N};-\frac{\gamma^2\tau^{2N}/\tau_1^{2N-2}}{p^2}\bigg) \Longrightarrow\  _2F_1\bigg(\frac{1}{2},\frac{1}{2},\frac{3}{2};-\frac{\gamma^2\tau^{2N}/\tau_1^{2N-2}}{p^2}\bigg)
\]
and
\[
_2F_1\bigg(\frac{1}{2},\frac{1}{2},\frac{3}{2};-\frac{\gamma^2\tau^{2N}/\tau_1^{2N-2}}{p^2}\bigg)= \frac{p\tau_1^{N-1}}{\gamma\tau^N}\ln(\gamma\tau^N/\tau_1^{N-1} + \sqrt{p^2+\gamma^2\tau^{2N}/\tau_1^{2N-2}})
\]

The approximate asymptotic behavior of $u_\pm(p,h,\tau)$ and $\upsilon_\pm(p,h,\tau)$ in the large $\tau$ limit is then found to be 
    \begin{equation}\label{eq:upm}
     u_\pm(p\hat{z},h,\tau)\simeq\frac{1}{\sqrt{2}}
      \exp\bigg[\mp i\chi\bigg]
\begin{bmatrix} 
\pm 1 \\
0 \\
+ 1 \\
0 \\
\end{bmatrix} ;\quad 
\upsilon_\pm(p\hat{z},h,\tau)\simeq\frac{i}{\sqrt{2}}
      \exp\bigg[\pm i\chi\bigg]
\begin{bmatrix} 
-1 \\
0 \\
\pm 1 \\
0 \\
\end{bmatrix}, 
    \end{equation}
    where, in the equation above,
\[
 \chi=\bigg(\frac{\gamma'}{N+1}\{\tau^{N+1}-\tau_0^{N+1}\}+\frac{p^2N}{\gamma'(N+1)}\bigg(\frac{1}{\tau^{N-1}}\ln{\tau^N}-\frac{1}{\tau_0^{N-1}}\ln{\pm\tau_0^N}\bigg)\bigg).
\]
For convenience, above we indicate with $\gamma'=\frac{\gamma}{\tau_1^{N-1}}$. Plugging this expression into the expressions above, we get the Bogoliubov coefficient $\beta(p)$ as
    \begin{equation}
        \beta(p)=-i\exp\bigg[-\frac{N\pi p^2}{\gamma(N+1)}\left(\frac{1}{\xi}\right)^{N-1}\bigg],
    \end{equation}
where we introduced the quantity $\xi\equiv\tau_0/\tau_1$. Notice that $\xi$ depends on the (in principle arbitrary, as long as preceding Big Bang Nucleosynthesis) conformal time at which the universe transitions to being radiation-dominated, and on the parameter $\tau_0$, which is chosen so that Eq.~(\ref{eq:upm}) is valid (see \cite{Boyle:2018rgh} for details on this).    Plugging this into (\ref{eq:beta}) gives the generalized number density
    \begin{equation}
    n_{dm}=\sum_{h}\int\frac{d^3p}{(2\pi)^3}\mid\beta_+(p)\mid^2=I \bigg(\frac{(N+1)\gamma}{2\pi N}\xi^{(N-1)}\bigg)^{3/2}.
\end{equation}
   In the equation above, $I$ is a dimensionless constant defined as 
   \[
I=\frac{1}{2\pi^2}\int_{0}^{\infty}x^2\bigg[1-\sqrt{1-e^{-x^2}}\bigg]dx   \approx 0.01276.
    \]
    Our result recovers the number density found in \cite{Boyle:2018rgh} for the radiation dominated background, $N=1$. To extract a prediction for the dark matter (right handed neutrino) mass out of this number density, we follow the same procedure as in \cite{Boyle:2018rgh}  (See their section 5.1.2).  With $n_{dm}$, we define the dark matter yield
\begin{equation}
    Y_{dm}=\frac{n_{dm}}{s},
\end{equation}
with $s$ the entropy density in the universe, which we assume to be conserved at early times and in particular at the transition to radiation domination.

After the reference conformal time $\tau_1$, we have the usual radiation-domination relations
\begin{gather}
    \rho=\frac{\pi^2}{30}g_*T^4, \qquad
    s=\frac{2\pi^2}{45}g_*T^3.
\end{gather}
From these relations, solving for temperature, one finds
\begin{equation}
    \frac{(2\rho)^{3/4}}{s}=\frac{3}{2}\bigg(\frac{15}{g_*\pi^2}\bigg)^{1/4}.
\end{equation}
    Combining all expressions for $n_{dm}$, $\gamma$ and $s$, we can express $Y_{dm}$ in terms of the particle mass $M_{dm}$

\begin{equation}
    Y_{dm}=\frac{3I}{2\pi^2}\bigg(\frac{15}{g_*}\bigg)^{1/4}\bigg(\frac{M_{\nu}}{\mu}\bigg)^{3/2}\bigg(\frac{N+1}{2N}\xi^{(N-1)}\bigg)^{3/2},
\end{equation}
    where $g_*=106.75$ is the number of the degrees of freedom in the standard model, and $\hat\mu=5.19\times10^8$GeV is the same right-handed neutrino reference mass scale  chosen by \cite{Boyle:2018rgh}. Notice that the formula, remarkably, does not depend on the conformal time $\tau_1$ at which radiation domination sets in (there could have in principle been a dependence both in the dark matter number density and via the energy density $\rho_1$).

    We can now match the mass of the dark matter particle and the number density to the present day dark matter energy density  $\rho_{dm}^{(0)}=9.7\times10^{-48}$GeV$^4$ and the present day entropy density $s^{(0)}=2.3\times10^{-38}$GeV$^3$

The late-time dark matter density is then given by    
    \begin{equation}
    \rho_{dm}^{(0)}=M_{dm}n_{dm}=M_{dm}Y_{dm}s^{(0)}.
\end{equation}
    Solving for the mass term for $M_{dm}=M_\nu$ gives the desired result for the sterile neutrino DM mass as function of $N$  The result is a sterile neutrino mass as a function of integer $N$ values.
\begin{equation}
    m_{\bar{\nu}} =
    4.8\times10^8\bigg(\frac{2N}{N+1}\bigg)^{3/5}\xi^{-\frac{3(N-1)}{5}}\textrm{GeV}\qquad ({\rm odd}\ N).
\end{equation}

\begin{figure}[t]
        \centering
        \includegraphics[width=0.6\textwidth]{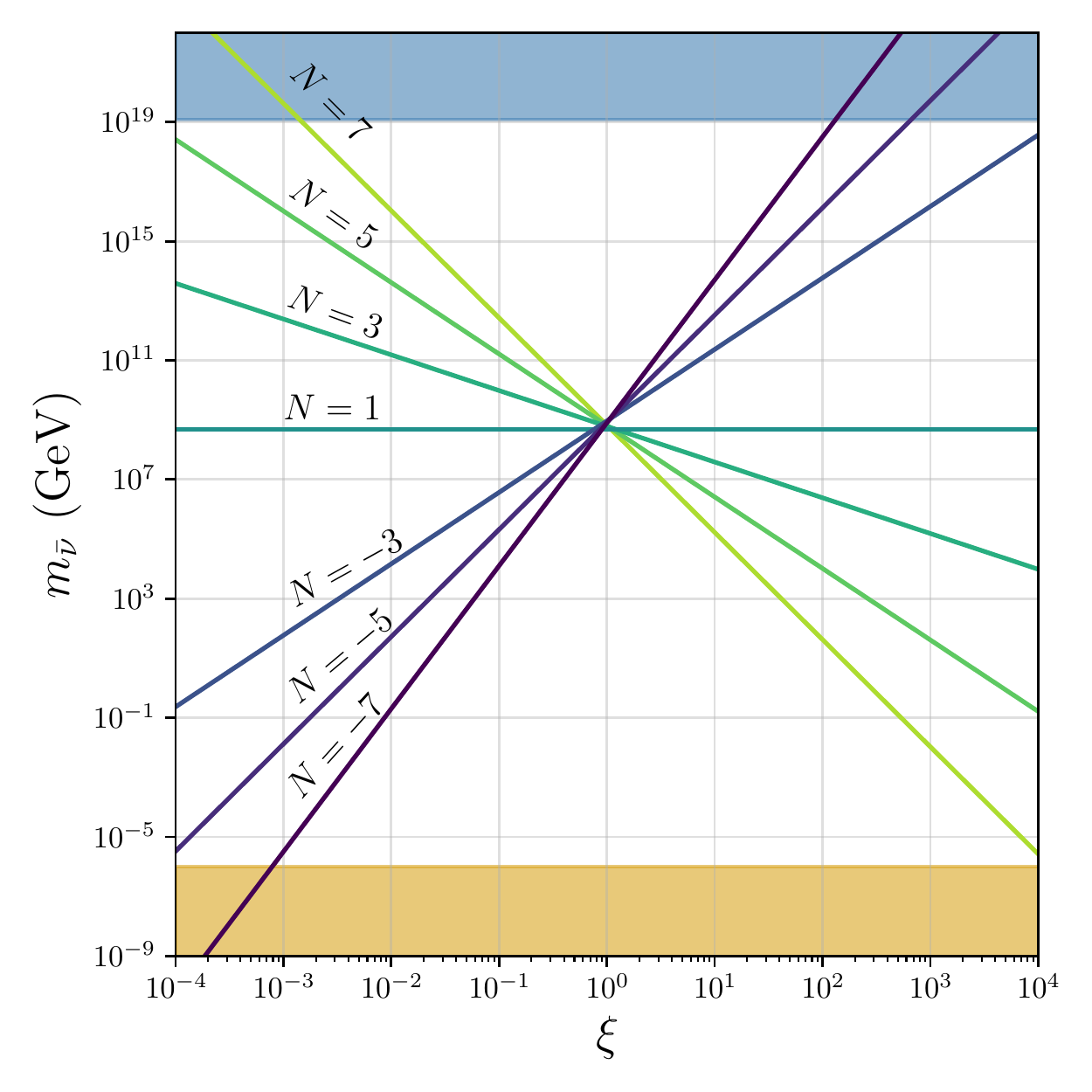}
        \caption{Mass of the right-handed neutrino giving the correct dark matter density as a function of $\xi$ for various values of $N$.}
        \label{fig:masses}
\end{figure}
  
 Fig.~\ref{fig:masses} shows, as a function of the parameter $\xi$, the mass of the right-handed neutrino $\bar\nu$ giving the right dark matter amount for a given value of the parameter $N$.
 
 This setup does not imply values for parameters such as the Majorana and Dirac masses of the neutrino sectors, and thus the lifetime of the dark matter particle is not a prediction of the model. We describe ways to detect sterile neutrino decay products in this scenario in the following section, in a model-independent way.

\section{Sterile Neutrino Decay Width}\label{sec:decay}

For simplicity, we assume the existence of a single, Weyl, sterile 
neutrino \(\vnur\). We take the sterile neutrino/SM interactions to stem from a 
Yukawa interaction between the RH neutrino and the SM lepton doublet. In this 
scenario, the Lagrangian density is given by:
\begin{align}
    \cL = \cL_{\mathrm{SM}} + 
    i\hat{\vnur}^{\dagger}\bar{\sigma}_{\mu}\partial^{\mu}\hat{\vnur} 
    -\frac{1}{2}\qty(\hat{m}\hat{\vnur}\hat{\vnur} + \mathrm{h.c.}) 
    + \qty(\epsilon^{ab}\vb{y}^{i}_{\nu}\phi_{a}\hat{L}_{bi}\hat{\vnur} + \mathrm{h.c.}).
\end{align}
In the expression above, \(\hat{m}\) is the Majorana mass of the RH neutrino, 
\(\vb{y}_{\nu}\) is a Yukawa vector with three components and \(\phi\), and 
\(\hat{L}\) are the Higgs and lepton doublets (note that all the above fermion fields are written in two-component notation). The doublets are given by:
\begin{align}
    \phi &= \mqty(G^{+} \\ \frac{1}{\sqrt{2}}\qty(v + h + i G^{0})), &
    \hat{L}_{i} &= \mqty(\hat{\nu}_{i} \\ \hat{\ell}_{i})
\end{align}
with \(G^{+}\) and \(G^{0}\) being the charged and neutral Goldstone bosons, 
\(h\) being the standard model Higgs, \(v \sim 246\) GeV being the Higgs 
vacuum expectation value, \(\hat{\nu}_{i}\) being the left-handed neutrinos 
and \(\hat{\ell}_{i}\) being the left-handed charged leptons. Expanding out 
the neutrino Lagrangian and gathering all the mass terms, we find the following:
\begin{align}
    -\cL_{\nu\mathrm{-mass}} &= \frac{1}{2}\hat{m}\hat{\vnur}\hat{\vnur} + 
    \frac{v}{\sqrt{2}}\vb{y}^{i}_{\nu}\hat{L}_{bi}\hat{\vnur} + \mathrm{h.c.}
    = \frac{1}{2}\vb{M}_{\nu}^{ij}\hat{\vb*{\nu}}_{i}\hat{\vb*{\nu}}_{j}
\end{align}
where the neutrino vector 
\(\hat{\vb*{\nu}} = \mqty(\hat{\nu}_1 & \hat{\nu}_2 &\hat{\nu}_3 &\hat{\bar{\nu}})^T\) 
and the \((4\times4)\) neutrino mass matrix is given by:
\begin{align}
    \vb{M}_{\nu} &= \mqty(0_{3\times3} & \frac{v}{\sqrt{2}}\vb{y}_{\nu} 
    \\ \frac{v}{\sqrt{2}}\vb{y}_{\nu}^T & \hat{m})
\end{align}
Diagonalizing the neutrino mass matrix is done through Takagi diagonalization 
with a unitary matrix \(\vb*{\Omega}\), satisfying 
\(\vb*{\Omega}^T\vb{M}_{\nu}\vb*{\Omega} 
= \mathrm{diag}(m_{\nu_{e}}, m_{\nu_{\mu}}, m_{\nu_{\tau}}, m_{\bar{\nu}})\). 
While it is possible to analytically compute \(\vb*{\Omega}\) for general
\(\vb{y}_{\nu}\), we choose to simplify our model by assuming only a single 
non-zero Yukawa coupling. That is, we take \(\vb{y}^{k}_{\nu} \equiv y\) and
\(\vb{y}_{\nu}^{k\neq0} = 0\) with \(k = 1, 2\) or \(3\). in this case, the
unitary Takagi matrix is given by an orthogonal \(\mathrm{O}(2)\) matrix 
times a diagonal unitary matrix:
\begin{align}
    \vb*{\Omega} &= \mqty(-i\cos\theta & \sin\theta\\ i\sin\theta &\cos\theta)
\end{align}
The angle satisfies
\begin{align}
    \cos\theta &= \frac{\sqrt{\hat{m}^2 + 2v^2y^2}+\hat{m}}{
        \sqrt{\qty(\hat{m} + \sqrt{\hat{m}^2 + 2v^2y^2})^2 + 2v^2y^2}
    } = \sqrt{\frac{m_{\bar{\nu}}}{m_{\nu}+m_{\bar{\nu}}}}\\
    \sin\theta &= \frac{\sqrt{\hat{m}^2 + 2v^2y^2}-\hat{m}}{
        \sqrt{\qty(\hat{m} - \sqrt{\hat{m}^2 + 2v^2y^2})^2 + 2v^2y^2}
    } = \sqrt{\frac{m_{\nu}}{m_{\nu}+m_{\bar{\nu}}}}
\end{align}
with the light and heavy neutrino masses \(m_{\nu}\) and \(m_{\bar{\nu}}\)
are given by:
\begin{align}
    m_{\nu} 
    &= \frac{1}{2}\qty(\sqrt{\hat{m}^2+2v^2y^2} - \hat{m}) 
    \sim \hat{m}\qty(\frac{v^2y^2}{2\hat{m}^2} + \order{\frac{vy}{\hat{m}}}^4)\\
    m_{\bar{\nu}} 
    &= \frac{1}{2}\qty(\sqrt{\hat{m}^2+2v^2y^2} + \hat{m})
    \sim \hat{m}\qty(1 + \frac{v^2y^2}{2\hat{m}^2} + \order{\frac{vy}{\hat{m}}}^4)
\end{align}
The mass eigenstates for the LH and RH neutrinos, \(\nu_k\) and \(\bar{\nu}\) 
are related to the interaction eigenstates via 
\(\vb*{\Omega}\mqty(\nu_k & \bar{\nu})^T\), or more explicitly:
\begin{align}
    \hat{\nu}_k &= -i\cos\theta \nu_k + \sin\theta \bar{\nu}\\
    \hat{\bar{\nu}} &= i\sin\theta \nu_k + \cos\theta \bar{\nu}
\end{align}
Using these transformations, we find that the Lagrangian density containing
the interactions between the RH neutrino and the standard model relavent to
this study is given by:
\begin{align}
    \cL 
    &= \cdots + \frac{g}{\sqrt{2}}\qty[
    i\cos\theta W^{+}_{\mu}\nu_{k}^{\dagger}\bar{\sigma}_{\mu}\ell_{k}
    +\sin\theta W^{+}_{\mu}\bar{\nu}^{\dagger}\bar{\sigma}_{\mu}\ell_{k}+\mathrm{h.c.}
    ]\\
    &\hspace{1.05cm} + \frac{e}{2s_{W}c_{W}}\qty[
    i\sin\theta\cos\theta Z_{\mu}\nu^{\dagger}_{k}\bar{\sigma}_{\mu}\bar{\nu}+\mathrm{h.c.}
    ]\notag\\
    &\hspace{1.05cm} +\frac{iy}{\sqrt{2}}h\nu_k\bar{\nu}\notag
\end{align}
where the \(\cdots\) are terms not containing interactions between the RH neutrino
and the SM or they contain irrelevant vertices (for example, interactions with
the Goldstones.) The partial decay widths for \(
\bar{\nu} \to W^{+} + \ell,W^{-} + \ell^{\dagger},Z + \nu\) and \(h + \nu\) are
given by:
\begin{align}
    \Gamma(N\to\nu_{\ell}+h) &= \dfrac{\lambda^{1/2}\qty(m_{\bar{\nu}}^2,m_{\nu}^2,m_{h}^2)}{16\pi m_{\bar{\nu}}v_H^2}\sin[2](\theta)\qty[(m_{\bar{\nu}}-m_{\nu})^2-m_{h}^2]\\
    %%%%%%%%%%%%%%%%%%%%%%%
    \Gamma(N\to W^{+}+\ell) &= \dfrac{e^2\sin[2](\theta)}{64\pi m_{\bar{\nu}}^3M_{W}^2s_W^2}\lambda^{1/2}\qty(m_{\bar{\nu}}^2,m_{\ell}^2,M_{W}^2)\\
                           &\qquad\times\qty[\qty(m_{\bar{\nu}}^2-m_{\ell}^2)^2+M_{W}^2\qty(m_{\bar{\nu}}^2+m_{\ell}^2)-2M_{W}^2]\notag\\
    %%%%%%%%%%%%%%%%%%%%%%%
    \Gamma(N\to Z+\nu_{\ell}) &= 
    \dfrac{e^2\sin[2](2\theta)}{256\pi m_{\bar{\nu}}^3M_{W}^2s_W^2}
    \lambda^{1/2}\qty(m_{\bar{\nu}}^2,m_{\nu}^2,M_{Z}^2)\\
                             &\qquad\times\qty(m_{\bar{\nu}}-M_Z-m_{\nu})
    \qty(m_{N}+M_Z-m_{\nu})
    \qty(2M_Z^2+\qty(m_{\bar{\nu}}+m_{\nu})^2)\notag
    %%%%%%%%%%%%%%%%%%%%%%%
\end{align}
where \(\lambda(a,b,c)=a^2+b^2+c^2-2ab-2ac-2bc\). Taking \(m_{\bar{\nu}} \ll m_{H}\), we
find:
\begin{align}
    \Gamma(N\to\nu_{\ell}+h) &\sim
    \dfrac{m_{\bar{\nu}}}{16\pi}\qty(\frac{m_{\bar{\nu}}}{v_{H}})^2\sin[2](\theta)\\
    %%%%%%%%%%%%%%%%%%%%%%%
    \Gamma(N\to W^{+}+\ell) &\sim
    \dfrac{e^2m_{\bar{\nu}}}{64\pi s_W^2}\qty(\frac{m_{\bar{\nu}}}{m_{W}})^2\sin[2](\theta)\\
    %%%%%%%%%%%%%%%%%%%%%%%
    \Gamma(N\to Z+\nu_{\ell}) &\sim 
    \dfrac{e^2 m_{\bar{\nu}}}{256\pi s_W^2}\qty(\frac{m_{\bar{\nu}}}{m_{W}})^2\sin[2](2\theta)
    %%%%%%%%%%%%%%%%%%%%%%%
\end{align}
The total decay with of the RH neutrino is roughly (for \(\theta \ll 1\)
\(m_{\bar{\nu}} \gg m_{H}\)):
\begin{align}
    \Gamma \sim \frac{10^{-6}}{\mathrm{GeV}^2}\theta^2 m_{\bar{\nu}}^3
\end{align}
and the corresponding lifetime
\begin{equation}
    \tau=\frac{6.6\times 10^{-28}\ {\rm s}}{\theta^2}\left(\frac{\rm 1000\ GeV}{m_{\bar\nu}}\right)^3,
\end{equation}
meaning that
\begin{equation}\label{eq:lifetime}
    \frac{\tau}{\tau_U}\simeq15\left(\frac{10^{-23}}{\theta}\right)^2\left(\frac{\rm 1000\ GeV}{m_{\bar\nu}}\right)^3,
\end{equation}
where we indicate with $\tau_U$ the age of the universe.

\section{Sterile Neutrino Thermalization}\label{sec:thermalization}
A necessary condition for the validity of the dark matter mass predictions discussed above is that the right handed neutrinos not thermalize: should they reach chemical equilibrium with the Standard Model thermal bath, the corresponding thermal relic density would be unrelated to the abundance from gravitational production. We must thus ensure that the parameter space under consideration be consistent with the absence of thermal equilibrium for the neutrino.

At $T>T_1$, where $T_1$ is the temperature at which the energy density of radiation equals the energy density that dominates in the early universe during gravitational dark matter production, 
$$
\rho\sim T^{2\frac{N+1}{N}}\qquad (T>T_1).
$$
Matching $\rho_{\rm rad}(T_1)$ with the expression above gives
$$
\rho\simeq \left(\frac{T}{T_1}\right)^{2\frac{N+1}{N}}T_1^4.\qquad (T>T_1).
$$
Right handed neutrinos can be in thermal equilibrium only when relativistic, which allows us to express the corresponding number density as $n_{\bar\nu}\sim T^3$ ($T>m_{\bar\nu})$. At large-enough temperatures, larger than both $m_{\bar\nu}$ and the electroweak scale, the cross section responsible for neutrino thermalization is approximately $\sigma_{\bar\nu-{\rm SM}}\sim\theta^2/T^2$, from decay and inverse decay into e.g. a gauge boson and a lepton.

In the radiation domination phase ($T<T_1$), when the Hubble rate is approximately $H\sim T^2/M_P$, thermalization happens at a temperature such that
\begin{equation}
    T^3\cdot\frac{\theta^2}{T^2}\sim \frac{T^2}{M_P}\quad \rightarrow\quad T\sim M_P\theta^2.
\end{equation}
Since by assumption $T<m_{\bar\nu}$, a sufficient condition to prevent thermalization is that $m_{\bar\nu}<M_P\theta^2$. From the discussion above, Eq.~(\ref{eq:lifetime}), for TeV and heavier right handed neutrinos, $M_P\theta^2\gg10^{-28}$ GeV, so thermalization can never occur during radiation domination.

In the non-standard phase, $T>T_1$, the Hubble rate is approximately
\begin{equation}
    H(T)\sim\frac{T_1^2}{M_P}\left(\frac{T}{T_1}\right)^{\frac{N+1}{N}}\ge \frac{T_1^2}{M_P}\qquad (T>T_1),
\end{equation}
thus to ensure that thermal equilibrium never be attained we simply need to require that 
\begin{equation}
    n_{\bar\nu}\sigma_{\bar\nu-{\rm SM}}\sim {T\theta^2}\ll\frac{T_1^2}{M_P}\le H(T)
\end{equation}
Below the Planck scale, $n_{\bar\nu}\sigma_{\bar\nu-{\rm SM}}< {M_P\theta^2}$, implying that equilibrium can only be attained for $T_1\sim \theta M_P$ which is much smaller than the scale of Big Bang Nucleosynthesis $T_{\rm BBN}\sim 1$ MeV, at which the universe must be radiation dominated (in other words, $T_1>T_{\rm BBN}$. Thus, thermal equilibrium is never attained at any point below the Planck scale.

\section{Constraints on Sterile Neutrino Lifetime}\label{sec:lifetime}

Tight constraints on the sterile neutrino lifetime can be computed using
experiments such as the Fermi Large Area Telescope (LAT) \cite{Atwood:2009ez}, IceCube \cite{aartsen2017constraints}, and HAWC \cite{Albert:2020ixl}. 

The gamma-ray and neutrino differential fluxes detectable on Earth from the 
decay of a sterile neutrino are given by:
\begin{align}
    \dv{\phi_{\gamma,\nu}}{E_{\gamma,\nu}} &= 
    \frac{1}{m_{\bar{\nu}}\tau}\dv{N_{\gamma,\nu}}{E_{\gamma,\nu}}
    \frac{J}{4\pi}
\end{align}
where \(\tau\) is the sterile neutrino life-time, 
\(\dd{N}_{\gamma,\nu}/\dd{E}_{\gamma,\nu}\) is the gamma-ray/neutrino spectrum
per decay and \(J\) is the so-called $J$-factor, the integral along the line of sight, and averaged on the telescope's angular aperture, of the dark matter density, which depends on the target being observed. Unlike for annihilation, the calculation of the $J$ factor for decay is prone to significantly less severe uncertainties from the slope of the inner density of the dark matter profile. Here, we use the same $J$ factors as in \cite{Coogan:2020tuf, Coogan:2021rez}.

\begin{figure}
    \centering
    \includegraphics[width=0.8\textwidth]{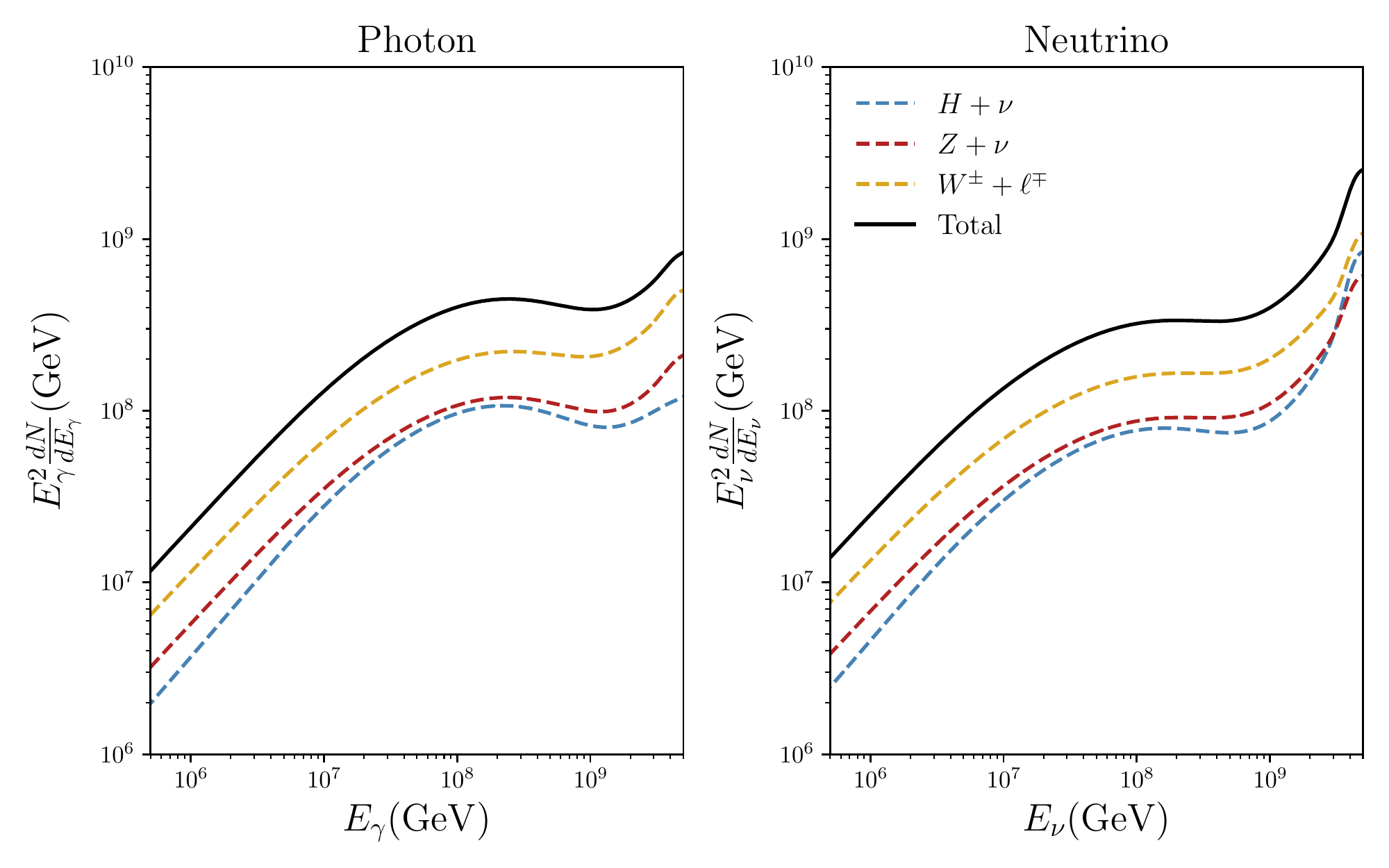}
    \caption{The gamma-ray (left) and neutrino spectra from the decay of a
    sterile neutrino with a mass of $4.8\times 10^8$ GeV (corresponding to the radiation domination case).}
    \label{fig:spectra}
\end{figure}

Computing the gamma-ray/neutrino spectrum for a sterile neutrino mass 
\(m_{\bar{\nu}} \sim 10^{8}\) GeV is a non-trivial task. Since 
\(m_{\bar{\nu}} \gg m_{W},m_{Z},m_{H}\), the electro-weak bosons emitted from
the sterile neutrino decays are highly-relativistic and are prone to undergo
large numbers of splittings into more electro-weak bosons. For this reason, we
use a newly developed software package, {\tt HDMSpectra}, specifically
developed for computing spectra from the decay of very heavy dark matter 
(\(m_{\mathrm{DM}}\gg m_{H}\)) \cite{bauer2020dark}. Fig.~\ref{fig:spectra} shows the gamma-ray
and neutrino spectra from the decay of a sterile neutrino with a mass of 
\(m_{\bar{\nu}}\sim4.8\times10^{8}\) GeV.

\begin{figure}
    \centering
    \includegraphics[width=0.5\textwidth]{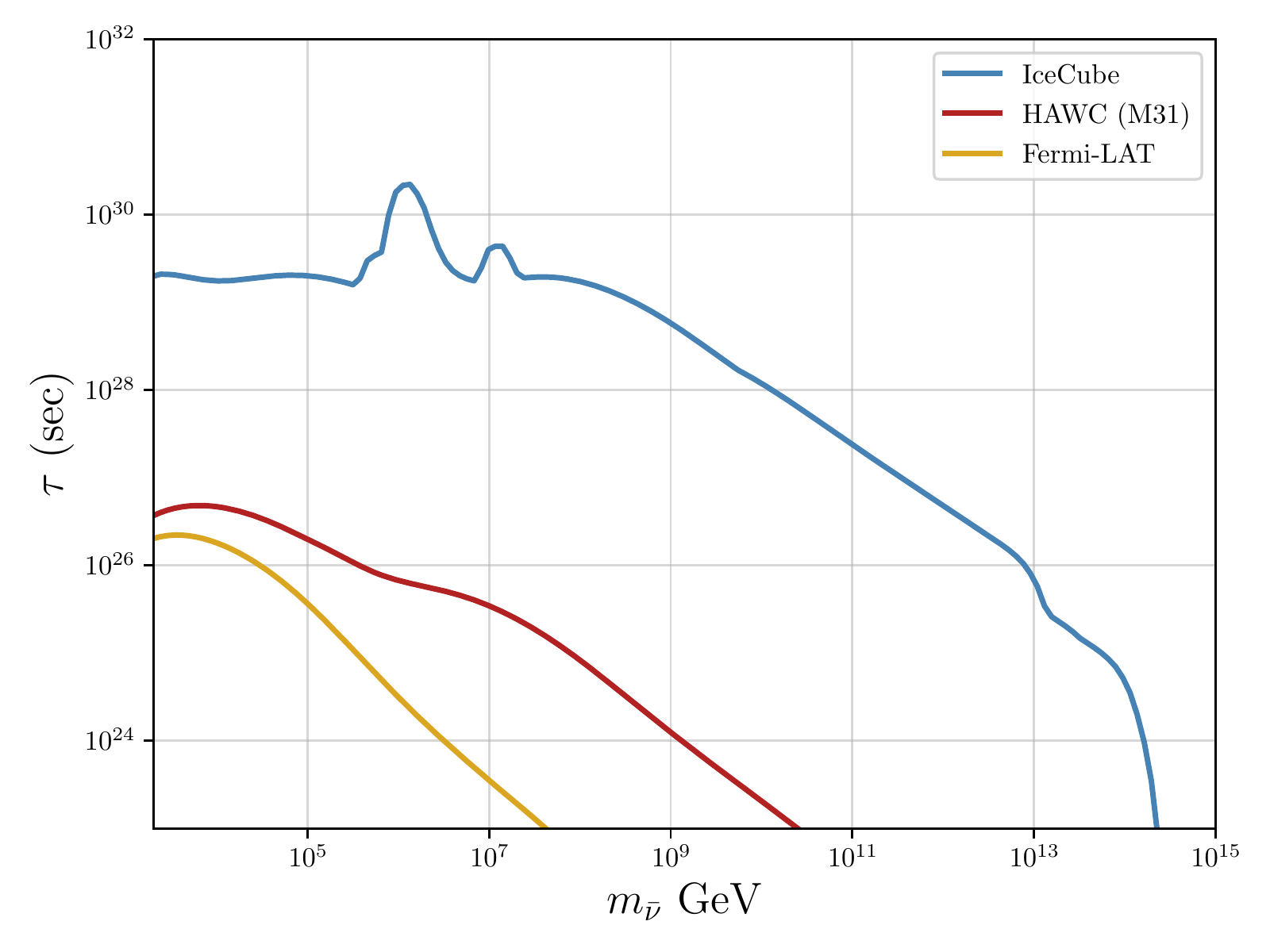}
    \caption{Upper limits on the right handed neutrino lifetime as a function of the particle mass, from IceCube \cite{aartsen2017constraints} (blue line), HAWC \cite{Albert:2020ixl} (red line), and Fermi LAT \cite{Atwood:2009ez} (yellow line).}
    \label{fig:constrants}
\end{figure}

To compute the constraints from HAWC data, we use the differential fluxes provided
in Ref.~\cite{Albert:2020ixl}. In that study, the HAWC collaboration measured
the gamma-ray fluxed from the Andromeda galaxy, an ideal target to look for dark matter decay debris. They measured fluxes of photons
with energies ranging from a TeV to 100 TeV and were able to place constraints
on \(E^2 \dd{N}_{\gamma}/\dd{E_{\gamma}} 
\lesssim 10^{-12}-10^{-11} \ \mathrm{TeV} \ \mathrm{cm}^{-2} \ \mathrm{s}^{-1}\).
Using these results, we find a constraint of the sterile neutrino lifetime of
roughly \(\tau \gtrsim 7\times10^{24} \ \mathrm{s}\) 
(see Fig.~(\ref{fig:constrants})).

To calculate constraints for  IceCube, we use the Galactic center neutrino upper limit fluxes shown in Fig.~2 of Ref.~\cite{aartsen2017constraints}; similarly, for the constraints from the Fermi Large Area Telescope (LAT) \cite{Atwood:2009ez}, we use the sensitivity discussed in  Ref.~\cite{Atwood:2009ez}.

We show the upper limits to the lifetime of a decaying right-handed neutrino in fig.~\ref{fig:constrants}. By far the best limits arise from the neutrino telescope IceCube, which in the mass range of typical interest, between $10^3$ and $10^8$ GeV, achieves a sensitivity in excess of $10^{29}$ s, two to three orders of magnitude better than gamma-ray telescopes. This lifetime sensitivity translates into probing mixing angles as small as $\theta\sim10^{-59}$.

\section{Summary, Discussion, and Conclusions}\label{sec:conclusions}
We generalized gravitational particle production in the early universe in $CPT$ symmetric cosmologies beyond the case of radiation domination discussed in Ref.~\cite{Boyle:2018rgh, Boyle:2018tzc}; our key result is that in non-standard, $CPT$ symmetric universes the mass of the dark matter candidate, a ``sterile'' right-handed neutrino, depends on the transition from the non-standard cosmological epoch to radiation, and on the dependence of the energy density on conformal time at early times. As a result, unlike the case of radiation domination, the dark matter mass in largely unconstrained and could range between the electroweak and the Planck scales. Note that we did not consider cases where the expansion history of the universe, while respecting $CPT$ invariance, changes at multiple points at early times: particle production in that case would happen during one particular phase, which would fall within the cases considered here.

We relaxed the assumption that the dark matter be protected from decay by a $Z_2$ symmetry, and discussed the possible detection of massive, right-handed neutrinos. We verified that the neutrino is never in thermal equilibrium in the early universe. We found that the optimal way to search for the neutrino's decay products is to employ high-energy neutrino telescopes such as IceCube, which at present rules out neutrinos with lifetimes in excess of $10^{29}$ s up to masses as large as $10^8$ GeV.

Future tests of this mechanism for the generation of the cosmological dark matter will depend upon observational indications for the lack of an inflationary period, as well as in favor of a $CPT$ symmetric early phase; the detection of the debris from the decay of a  superheavy species compatible with that species being a right-handed neutrino would also be a smoking-gun signature for this scenario.

\acknowledgments

LM and SP are partly supported  by the U.S. Department of Energy grant number DE-SC0010107.

\bibliography{references}

\end{document}